\documentclass[a4paper]{article}
\usepackage{cite}
\usepackage{amsmath,amssymb,amsfonts}
\usepackage{algorithmic}
\usepackage{graphicx}
\usepackage{textcomp}
\usepackage{bmpsize}
\usepackage{xcolor}
\usepackage{lipsum}
\usepackage[colorlinks=true,urlcolor=black]{hyperref}
\usepackage{authblk}
\usepackage{hyperref}
\usepackage{amsfonts}
\usepackage{amsmath}
\usepackage{graphicx}
\usepackage[space]{grffile}
\usepackage[margin=2.5cm]{geometry}
\usepackage{pdfpages}
\usepackage[capitalize,noabbrev]{cleveref}

\usepackage[]{algorithm2e}
\usepackage{todonotes}

\providecommand{\keywords}[1]{\textbf{\textit{Keywords ---}} #1}
\graphicspath{{img/}}

\def\BibTeX{{\rm B\kern-.05em{\sc i\kern-.025em b}\kern-.08em
    T\kern-.1667em\lower.7ex\hbox{E}\kern-.125emX}}
\begin{document}

\title{Probing Channel Balances in the Lightning Network}

\iftrue
\author[1]{Sergei Tikhomirov \thanks{sergey.s.tikhomirov@gmail.com}}
\author[2]{Rene Pickhardt \thanks{rene.m.pickhardt@ntnu.no}}
\author[1]{Alex Biryukov \thanks{alex.biryukov@uni.lu}}
\author[2]{Mariusz Nowostawski \thanks{mariusz.nowostawski@ntnu.no}}
\affil[1]{University of Luxembourg \\
Esch-sur-Alzette, Luxembourg}
\affil[2]{Department of Computer Science \\
Norwegian University of Science and Technology\\
Gj{\o}vik, Norway}



\maketitle

\begin{abstract}
Payment channel networks (PCN) have been introduced to solve scalability problem of permissionless blockchains such as Bitcoin.
The Lightning Network (LN), launched in 2018, is the most prominent PCN built on top of Bitcoin.
As of March~2020, LN consists of over 11~thousand nodes and 36~thousand channels, collectively holding nearly $900$~bitcoins ($7.8M$~USD).

A payment channel allows two parties to lock funds in a multisignature address and then modify the distribution of funds in nearly instant transactions, without confirming them in the blockchain.
As LN transactions are not broadcast and not publicly stored, LN has been seen not only as scalability but also as privacy solution.
The protocol guarantees that only the latest channel state can be confirmed on channel closure.
LN also supports multi-hop payments, where the balances of multiple channels along a path are moved.
The atomicity of this process is guaranteed by hash time-locked contracts (HTLC).
LN nodes gossip about channels available for routing and their total capacities.
To issue a (multi-hop) payment, the sender creates a route based on its local knowledge of the graph.
As local channel balances are not public, payments often fail due to insufficient balance at an intermediary hop.
In that case, the payment is attempted along multiple routes until it succeeds.
This constitutes a privacy-efficiency tradeoff: hidden balances improve privacy but hinder routing efficiency.

In this work, we show that an attacker can easily discover channel balances using probing.
This takes under a minute per channel and requires moderate capital commitment and no expenditures (the attacker's funds are only temporarily locked).
We describe the algorithm and test our proof-of-concept implementation on Bitcoin's testnet.
We argue that LN's balance between privacy and routing efficiency is suboptimal: channel balances are neither well protected nor utilized.

We outline two ways for LN to evolve in respect to this issue.
To emphasize privacy, we propose a modification of error handling that hides details of the erring channel from the sending node.
This would break our probing technique but make routing failures more common, as the sender would not know which channel from the attempted route has failed.
To improve efficiency, we propose a new API call that would let the sender query balances of channels that it is not a party of.
We argue that combining these approaches can help LN take the best of both worlds: hide private data when feasible, and utilize public data for higher routing efficiency.
\end{abstract}


\twocolumn

\section{Introduction} \label{sec:introduction}
Bitcoin~\cite{nakamoto2008bitcoin} is the first digital currency to solve the double spending problem without a trusted third party.
The permissionless nature of Bitcoin's consensus algorithm imposes severe restrictions on transaction throughput, with a theoretical maximum of tens of transactions per second~\cite{croman2016scaling}.
Addressing this problem in a naive way (such as increasing the block size) would endanger Bitcoin's key properties, such as the ability of participants to validate all transactions in real time using consumer hardware.
Another scaling approach that doesn't affect Bitcoin's properties but builds on top of them is \textit{off-chain scaling}.
Off-chain, or \textit{second layer} (\textit{L2}) protocols allow parties to exchange transactions without broadcasting them to the blockchain, but in case of a conflict the blockchain (called \textit{layer one} in this context) is used for dispute resolution.
We refer the reader to a comprehensive overview of off-chain approaches in~\cite{gudgeon2019sok}.

The (Lightning Network) LN has gained significant traction as a major Bitcoin scaling solution.
First introduced in 2016~\cite{poon2016bitcoin} and launhed on mainnet in early~2018, it is being implemented by several independent teams based on a common set of specifications called BOLTs (for Basics of Lightning Technology).

The basic building block of LN is a \textit{payment channel}.
A payment channel is a protocol for off-chain transactions based on the initial \textit{funding transaction}, which locks coins in a 2-of-2 multisignature address.
A channel operates in three stages: opening (two parties lock the coins), operating (exchanging off-chain transactions), and closing (broadcasting the most recent channel state to the blockchain).
A channel partner may try to cheat by trying to close the channel with an earlier state.
In this case, the counterparty is guaranteed to get all funds from the channel if they react within a timeout.

LN nodes gossip about newly opened channels that are marked as available for routing.
Based on this information, each node maintains a local view of the network, and uses it to calculate routes to the receiver.
As total channel capacities are known, the sender only considers channels with the capacity larger than $X$ for a payment of amount $X$.
However, this is insufficient to prevent routing failures, because the distribution of funds in channels is not known.
The ability of channel parties to send or forward payments is limited by their \textit{local} channel balances.
Initially, if Alice opens a channel with Bob, all funds are on her side.
This means that she can send up to the total capacity, but she can not receive.
As the local balances change, the routing capabilities of the channels in both directions are also changing.

In case of payment failure, the sender is notified by an error message which channel has failed, and will presumably re-launch route generation function with the failed channel excluded.
This procedure repeats for a pre-set number of tries, or until a payment succeeds.
Given the external constraints it is sill not clear that a better approach than systematically probing for paths exist even though active research is being conducted in this area~\cite{pickhardt2019imbalance, prihodko2016flare, grunspan2018ant, pickhardt2019jit, piatkivskyi2018split, sivaraman2018routing, bagaria2019boomerang,roos2017settling}.

Another aspect of LN relevant for this work is privacy.
Bitcoin is pseudonymous: while blockchain addresses are not linked to real world identity, and users are encouraged to not reuse them, the transaction graph can be extracted and analyzed.
In contrast, LN payments are only sent to a short chain of randomly chosen nodes along the route from a sender to a receiver.
Due to onion routing, each intermediary node only knows the previous and the next hops, but not the whole route and even not its position in the route.
This supports the presumption that Lightning payments provide more privacy than layer-one Bitcoin transactions.
However, a probing attack described in~\cite{van2019improvements} showed that it is possible for an attacker to reveal balances of other channels of their channel partner.

We see an important tradeoff in LN between routing efficiency and privacy.
On the one hand, hiding balances from everyone but channel parties is an important privacy measure.
If an external observer could probe intermediate channel balances, it would be possible to track payments from the sender to the receiver, thereby breaking relationship anonymity and value privacy (as defined e.g.~in~\cite{malavolta2017concurrency}).
On the other hand, knowing channel balances would allow senders to avoid trying routes that will fail.

We argue that LN in the current form takes "the worst of both worlds": information about channel balances is neither well protected nor utilized.
We support the first part of this claim by describing a probing algorithm that reveals channel balances by sending fake payments.
Contrary to~\cite{van2019improvements}, which requires opening channels with both parties of the channel that is being probed, our approach allows to probe the whole network while only establishing a handful of channels.
This greatly reduces the cost of the attack.
We test our proof-of-concept implementation on Bitcoin testnet and successfully probe a significant portion of channels with high precision.
We also outline potential countermeasures.
We leave the formal treatment of the second part of the claim (how utilizing balance information would improve routing) for future work.

The rest of the paper is organized as follows.
In~\cref{sec:background}, we describe the relevant aspects of LN in more detail.
In~\cref{sec:probing}, we introduce our probing algorithm.
In~\cref{sec:results}, we describe the experimental results.
We discuss the results and propose countermeasures in~\cref{sec:discussion}.
\cref{sec:related-work} provides a short summary of related work, and~\cref{sec:conclusion} concludes.

\section{Background} \label{sec:background}
In this section, we introduce the basic concepts of the LN and the details relevant for our work.

Each \textit{LN node} is defined by a private-public key pair.
A node has a permanent identifier and connects to other nodes in the LN P2P network.
Nodes exchange information about the currently open channels and their fee policies.
Nodes communicate with an underlying bitcoin node (such as Bitcoin Core) to receive information on which transactions are confirmed.

To create an \textit{LN channel}, Alice issues a request to Bob via the P2P protocol.
If the parties agree on channel parameters, they co-sign a \textit{funding transaction} that establishes the initial distribution of funds.
While in the initial specifications it was assumed that both parties could fund a channel, the current LN channels are single-funded (i.e.,~Alice provides all funds and may optionally "push" some funds to Bob as a gift).
After the funding transaction gets the sufficient number of confirmations (usually~$3$~to~$6$), the channel is open.

An \textit{LN transaction} is an atomic update of one or multiple channels.
To send a payment to Bob, Alice negotiates a new channel state.
Each channel state is reflected in a \textit{commitment transaction} -- a bitcoin transaction that spends the funding transaction and distributes the funds to both parties in some proportion.
Outputs of commitment transactions are called \textit{Hash time-locked contracts}, or \textit{HTLC}s.
HTLC is defined by a Bitcoin script that gives the funds either to one party, if it provides a pre-image of a given hash, or to the other party after a timeout.
For instance, if Alice wants to send $x$ coins to Bob, she first asks him for a \textit{payment hash}.
Bob generates a random number $r$ and send its hash $H(r)$ to Alice in a message called an \textit{invoice}.
Alice then \textit{offers} Bob an HTLC that would give him $x$~coins if he provides a preimage of $H(r)$ before time $t$, or return the coins to Alice otherwise.
Bob must \textit{claim} the payment before time $t$.
If he provides $r$, the HTLC \textit{resolves}: the balances in the next commitment transaction outputs will reflect the new distribution of channel funds.
The other way for an HTLC to be resolved is timeout.
A payment channel can keep track of multiple unresolved HTLCs concurrently.

LN is source routed.
Nodes exchange information about channels, and each node chooses routes based on the local view of the network graph.
The graph includes total channel capacities but not local balances of the counterparties.
Using HTLCs enables atomicity of channel updates in multi-path transactions.
For instance, Alice can pay Charlie via Bob by using the same hash value along the route.
If Charlie redeems the payment from Bob, then Bob can use the same pre-image to claim funds from Alice, otherwise no funds move.

The cornerstone of LN security is its \textit{revocation} mechanism.
The party that initiates the channel closure must wait until their portion of the funds becomes available.
For instance, if Alice initiates a channel closure, she must wait until her funds become available.
If the closure is a cheating attempt, Bob can dispute it before the timeout and take control of all channel funds.

Multi-path payments use onion routing to enforce the order of intermediary nodes.
Each HTLC is onion-encrypted so that each intermediary node only knows the immediate previous and next nodes, but not the final sender or receiver.
Transaction fees are collected by intermediary nodes as a difference between the amount in the HTLC they receive and the HTLC they send as part of the same transaction.
Note that if a payment fails, no fees are collected, as all pending balance updates roll back.


\section{Probing algorithm} \label{sec:probing}
We describe our algorithm of channel probing.

Let us denote the total capacity of a channel as $c$.
The two parties of a channel are denoted as source (with balance $b_s$) and destination ($b_d$).
As per BOLT specification, the source is defined as the node with an alphanumerically smaller node ID.
For concreteness, for each channel, we estimate the balance at source ($b_s$), and refer to it simply as $b$.

First, we establish the list of channels we are going to probe.
To be suitable for probing, the channel must be \textit{active} (available for routing) and \textit{live} (responding to requests).
As a P2P network, LN has no definitive list of live nodes.
Therefore, we need heuristics to choose live channels for probing. 
Probing dead channels would be wasteful: we would have to wait for timeouts, and the probes would leave our channels with "hanging" HTLCs waiting to be redeemed, occupying our capacity.

We define a channel as active if it is marked as active in at least one of the two directions (this information is available in gossip data).
To determine liveness, we use two heuristics: temporarily connecting to all nodes, and sending a small amount to all active channels (pre-probing).
After compiling a list of live and active channels, we probe them by sending payments with randomly generated payment hashes.
We refer to such payments as \textit{probes}.
For each channel, we keep a lower ($b_{min}$) and an upper ($b_{max}$) bound for its balance $b$.
Initially, $b_{min}=0$ and $b_{max}=c$.
At each probing step, we aim at shrinking this interval by issuing a probe with the amount of $0.5 * (b_{min} + b_{max})$ (binary search).\footnote{If the midpoint between $b_{min}$ and $b_{max}$ is larger than the maximum HTLC amount allowed by the specification, we decrease it to that maximum minus a safety margin.}

The two directions of a channel can have different properties.
Each channel partner controls its channel direction: Alice sets the routing policy only in the direction to Bob, while Bob sets the policy only in the direction to Alice.
It is possible that a channel is only partially active: Alice allows routing to Bob, but Bob does not allow routing to Alice.
Therefore, we try probing each channel in both directions, taking advantage of the fact that the two local balances sum up to the channel capacity: $b_s + b_d = c$.
This helps us probe partially active channels.
It is also helpful for probing large channels: if we do not have enough capacity to probe a channel from the "large" end, we probe it from the "small" end, getting the same information.
For clarity, we omit this implementation detail from the algorithm description.

We now describe our probing technique more thoroughly.

\subsection{Selecting channels for probing}
We use the following heuristics to determine liveness (Algorithm~\ref{alg:select-channels}).

\subsubsection{Heuristic 1: Connecting to nodes}
For a channel to be live, both its parties must be live.
We extract a list of nodes from gossip data and establish a P2P connection to each.\footnote{Establishing a P2P connection is nearly instant and does not require coins, unlike opening a channel.}
We consider a channel live, if both its parties are live.

\subsubsection{Heuristic 2: Pre-probing}
To further optimize probing, we introduce an additional pre-probing step.
We send a probe of $1000$~satoshis to every channel marked as active in the gossip data.
If we get an error of one of the two types that we use for probing, we consider the channel live, otherwise we consider it dead an do not consider it it in the main probing step.
We start the first main round of probing with a list of live channels where a channel is defined live if either heuristic 1 or heuristic 2 detected it as live.

\subsubsection{Heuristic 3: Liveness detected during probing}
If we issue a probe along the route of channels $c_1, c_2, \dots, c_n$ and receive an error from channel $c_i$, we conclude that all channels $c_j, j<=i$ are live.
If any of $c_j$ is not on our live channels list, we add it.
During the second probing round, we use the updated live channels list.

\begin{algorithm}
	\KwData{Gossip data}
	\KwResult{Channels selected for probing}
	\For{node in gossip data} {
		connect to node\;
		\If{connection established}{add node to live nodes\;}
	}
	\For{channel in gossip data}{
		\If{source and destination in live nodes}{
			add channel to channels to probe\;
		}
	}
	\For{channel in gossip data} {
		send a 1000~sat probe\;
		\If{error returned}{add channel to channels to probe\;}
	}
	\caption{Select channels for probing}
	\label{alg:select-channels}
\end{algorithm}

\subsubsection{Channel order}
Our method is agnostic to the order in which the channels are probed.
However we probe the "closer" and "more important" channels first.
The rationale behind this is that knowing the balances of channels often used as intermediary hops allows us to avoid sending payments that are known to fail early due to insufficient balance at some hop (note that the standard routing does not know about the funds distribution within channels).
We refer to the nodes we open channels with as \textit{entry nodes}.
We probe channels in the following order:

\begin{itemize}
	\item the "first layer" -- channels adjacent to the entry nodes;
	\item channels between hubs -- channels connecting nodes out of 1\% of the most connected nodes (if not already probed);
	\item the "second layer" -- channels adjacent to the "first layer" (if not already probed);
	\item other channels (if not already probed).
\end{itemize}

\subsection{Probing}

\subsubsection{Probing a single channel}
The key idea behind our method is that by sending a payment of amount $a$ to a channel, we can use the type or error it returns to infer whether its balance is higher or lower than $a$.
We use a randomly generated value as a payment hash, therefore, the payment fails in any case.
Let $c_1, c_2, \dots, c_n$ be a route we use, and $b_i$ be their respective balances.
Let $c_j$ be the erring channel.
After each probe with an amount $a$, we obtain the following information: 
\begin{itemize}
	\item $b_i > a$ for $i<j$;
	\item $b_j < a$ if the error is "insufficient capacity", or $b_j > a$ if the error is "unknown preimage"\footnote{The latter is only possible if $j=n$.}.
\end{itemize}

The probing algorithm for a single route is presented in Algorithm~\ref{alg:probe-route}.

\begin{algorithm}
	\KwData{Route and amount to probe}
	\KwResult{Updated balance estimates for channels in route}
	send payment along route\;
	\For{channels before erring channel} {
		$b_{min} = a$\;
	}
	\For{erring channel}{
		\If{insufficient funds}{
			$b_{max} = a$\;
		}
		\If{unknown preimage}{
			$b_{min} = a$\;
		}
	}
	\caption{Update balance estimates for channels along a route}
	\label{alg:probe-route}
\end{algorithm}

The algorithm for all channels selected for probing is presented in Algorithm~\ref{alg:probe-channels}.

\begin{algorithm}
	\KwData{Gossip data}
	\KwResult{Improved estimates for channels}
	get channels for probing\;
	\For{channel in channels for probing} {	
	    $b_{min} = 0$\;
	    $b_{max} = c$\;
		\For{number of probings per channel} {
			\For{number of attempts per probing}{
				get route to target\;
				probe route\;
				\If{target channel estimates updated}{
					continue\;
				}
			}
			\If{required precision reached}{
					continue\;
				}
		}
	}
\caption{Probe all channels}
	\label{alg:probe-channels}
\end{algorithm}

\subsubsection{Choosing routes}
We rely on the built-in functionality of our LN node to calculate routes to target channels.
Additionally, we pre-filter suggested routes based on the information we have obtained through probing.
If we know that a balance of some channel in a suggested route is insufficient, we exclude such channel from consideration for the current probe and repeat route search (Algorithm~\ref{alg:find-route}).

Optimizing route length presents a trade-off.
Choosing longer routes allows to collect more information per probe, but we are less likely to reach the target channel, as the probability that some intermediary channel fails is higher.
For our purposes, we limit the length of routes to $10$~hops (the protocol limit is $20$~hops).

\begin{algorithm}
	\KwData{target channel, amount $a$}
	\KwResult{Route to target suitable for $a$}
	\For{channels adjacent to destination} {
		\If{channel is not target}{
			add channel to excluded channels\;
		}
	}
	\For{all channels}{
		\If{$a > c_{max}$}{
			add channel to excluded channels\;
		}
	}
	\While{route is bad}{
		get route to target without excluded channels\;
		\For{channel in route}{
			\If{$a > b_{max}$}{
				route is bad\;
			}
		}
	route is good\;
	}
	\Return route\;
	\caption{Find a suitable route to target channel and given amount}
	\label{alg:find-route}
\end{algorithm}

\subsubsection{Second probing pass}
We discovered that some channels we thought were dead are in fact live, as our payments were successfully routed through them (liveness heuristic 3 described earlier).
We mark such channels as live and repeat the probing the second time taking them into consideration.

\subsection{Channel information coefficient}
To measure the effectiveness of our technique, we introduce the channel information coefficient.
For a given channel, it is defined as:

\[i = 1 - \frac{b_{max} - b_{min}}{c}\]

where $c$ is the original channel capacity, and $b_{max}$ and $b_{min}$ are our upper and lower bound estimates for $b$.
Thus $i=0$ would mean that we do not know any extra information about $b$ compared to public knowledge.
The value $i=1$ means that we know $b$ precisely.
In practice, if no unexpected error occurs, after $7$ probings we shrink the interval between the minimum and the maximum estimates to $1/128$ of the initial estimates, that is, we determine the channel balance with the precision better than $1\%$.

\subsection{Implementation details}

We implemented the algorithm described in~\cref{sec:probing} as a c-lightning plugin.
C-lightning~\cite{clightning} is one of the three major LN implementations, alongside LND~\cite{LND} and Eclair~\cite{Eclair}.
The plugin functionality allows developers to integrate their code in Python with the c-lightning node~\cite{clightningPlugins}.
We used five~entry channels opened to handpicked nodes with high connectivity and liquidity.
Four of our channels had the maximum allowed capacity of $0.167$~BTC, and one channel had the capacity of $0.043$~BTC.

\subsubsection*{Ethical considerations}
For ethical reasons and to be in compliance with privacy-related regulations, we only collected data from Bitcoin~testnet.

\section{Results} \label{sec:results}
\subsection{State of Bitcoin testnet}

At the time of the experiment (2020-02-26), Bitcoin~testnet contained $1974$ nodes and $5884$ channels (including $2527$ announced as active).
The initial estimate showed that $207$ nodes and $1625$ channels were live.
We detected $3$ additional live channels during the main probing phase.
The strongly connected component of the live subgraph contained $1489$ channels.
The other $139$ channels where pointing towards nodes for which we could not get meaningful error messages.


\subsection{Conducting the probing}

We launched the code that implements the algorithm described in~\cref{sec:probing}.

In total we sent out $12895$ onions: $3153$ ($24.45\%$) during the pre-probing and $9742$ ($75.55\%$) during the main probing).
Out of $9742$ onions in the main probing, $8256$ ($84.75\%$) returned errors which we could use to improve the estimates of balances on payment channels (namely, "channel temporary unavailable", which we interpret as "insufficient balance", and "payment details unknown", which we interpret as "preimage unknown").
Our code was running for~$14$ hours and~$6$~minutes. 
Probing $1628$ live channels took $65\%$ of the time (roughly $9$ hours), while the rest of the time was spent on slow responding channels or channels which replied with an unexpected error.

\subsection{Probing parameters}
One of the decisions we had to make is to set a timeout after which we declare a channel unresponsive and move to the next one.
The LN protocol does not prescribe how quickly a node should react.
The only time limitation is the HTLC timeout, which is usually on the order of hours or days.
Therefore, to probe all channels in reasonable time, we has to set a timeout on the order of seconds.
For the main probing, we choose a $10$~seconds timeout.
Our later results show that this is a reasonable tradeoff between probing speed and accuracy.

\subsection{Distribution of probing times}
Figure~\ref{fig:route_length_timings} shows the distribution of probing times for various route lengths.
Nearly all onions of $1$, $2$ or $3$ hops return within $10$ seconds.
The median of the $8256$ onions is at $3.36$ seconds.
Recalling that $15.25\%$ of the onions timed out or gave a non usable error back, we computed the corrected median without the timed out and erring onions, which is $3.93$ seconds. 
We conclude that the cutoff at $10$ seconds is an acceptable tradeoff.
Note that the diameter of the strongly connected component on the LN is just $3$.\footnote{Note however, that it is often impossible to use the shortest route, as it may not have sufficient balances in all channels on the side of the sender.}

\begin{figure}[]
  \centering
  \includegraphics[scale=0.45]{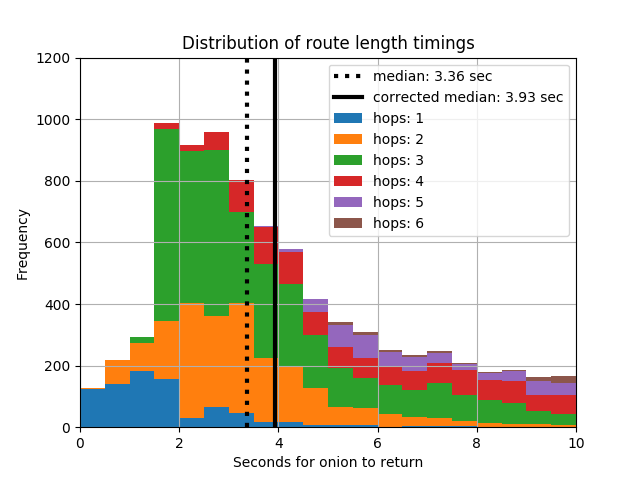}
  \caption{Distribution of probing timings on an onion level also drilled down depending on the distance in the route where the onion failed.}
  \label{fig:route_length_timings}
\end{figure}


\subsection{Time to probe a channel}

We look at the cumulative distribution function of times spent on channels in Figure~\ref{fig:channel_timings_CDF}.

\begin{figure}[]
  \centering
  \includegraphics[scale=0.45]{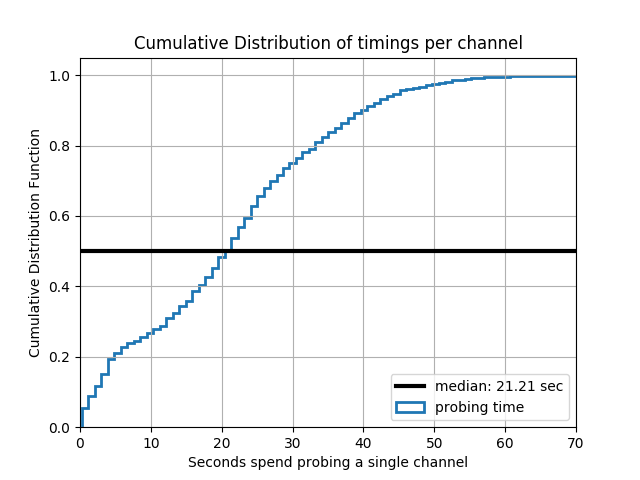}
  \caption{Distribution of probing timings on a channel level}
  \label{fig:channel_timings_CDF}
\end{figure}

All channels can be probed within $1$ minute.
This is very plausible considering the timeout of $10$ seconds for onions and the fact that we did only $7$ steps in binary search, so the expected worst case probing time is $70$ seconds.
However it is very unlikely that we have $7$ onions which are all short below $10$ seconds.
We furthermore observe that $50\%$ of channels can be probed in less than $21.2$~seconds.

\subsection{Probing results}

We used the channel information coefficient as introduced in Figure~\ref{fig:cdf_channel_coefficients} to measure how much information we obtained for all the channels that we probed.
We separated the channels into small and large channels.
We labeled as small channels those with capacity less then two times the maximum HTLC value (i.e. up to $2*0.043=0.086$~BTC). Small channels could be at least theoretically fully probed.

\begin{figure}[]
  \centering
  \includegraphics[scale=0.45]{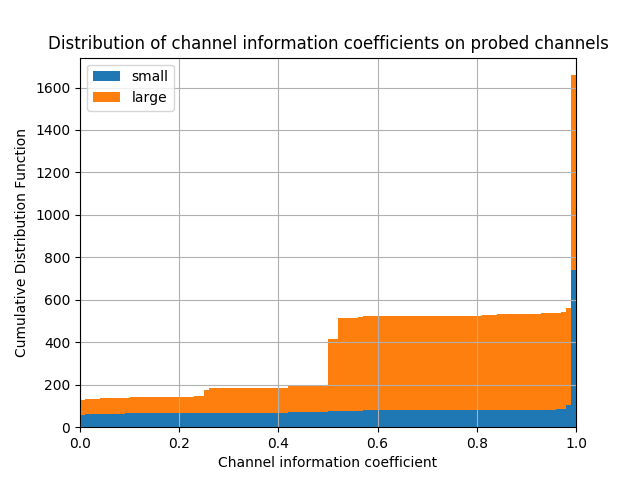}
  \caption{Distribution how often channel information coefficients shows that we can reveal full balance information for a majority of the attacked channels.}
  \label{fig:cdf_channel_coefficients}
\end{figure}

We obtained full information about the channel balance on over~$1000$ of~$1628$ channels.
Also we note the jump at $0.5$~for the large channels.
This is expected, as large channels have roughly $4$~times the capacity of the maximum HTLC amount limit, therefore if we probe $25\%$~from each side, this leaves us with a channel information coefficient of $0.5$~for the channels that have more than $25\%$~of the capacity on either side.

The notion of channels being \textit{balanced} is important for LN.
Intuitively, a channel is \textit{balanced} if it has roughly comparable balances on either side.
If the two balances differ significantly, the channel is \textit{unbalanced}.
We introduce the \textit{balance coefficient} to quantify this.
The balance coefficient represents the distance from the actual channel balance to $0.5$~of the total capacity, where $b$ is the estimated local balance and $c$ is the total channel capacity:

\[c_{bal} = 0.5 - \frac{|b-c|}{c} \]

$c_{bal} = 0$ means the channel is unbalanced: all balance is on one side.
$c_{bal} = 1$ means the channel is perfectly balanced: there are equal local balances on both sides.
Figure~\ref{fig:balance-coeff-histogram} depicts the distribution of balance coefficients among "small" channels that we were able to probe with high accuracy (information coefficient $> 0.9$).
We do not include large channels in this figure, because for many of them we can not precisely estimate the balance, hence they would appear perfectly balanced in the figure.
We conclude that many channels are indeed unbalanced: $15\%$~have the balance coefficient lower than $0.001$, $45\%$~lower than $0.01$, and $62\%$~lower than $0.1$.
Note, however, that the picture may change if we consider large channels, and that channel management practices on mainnet may differ significantly.

\begin{figure}[]
  \centering
  \includegraphics[width=0.4\textwidth]{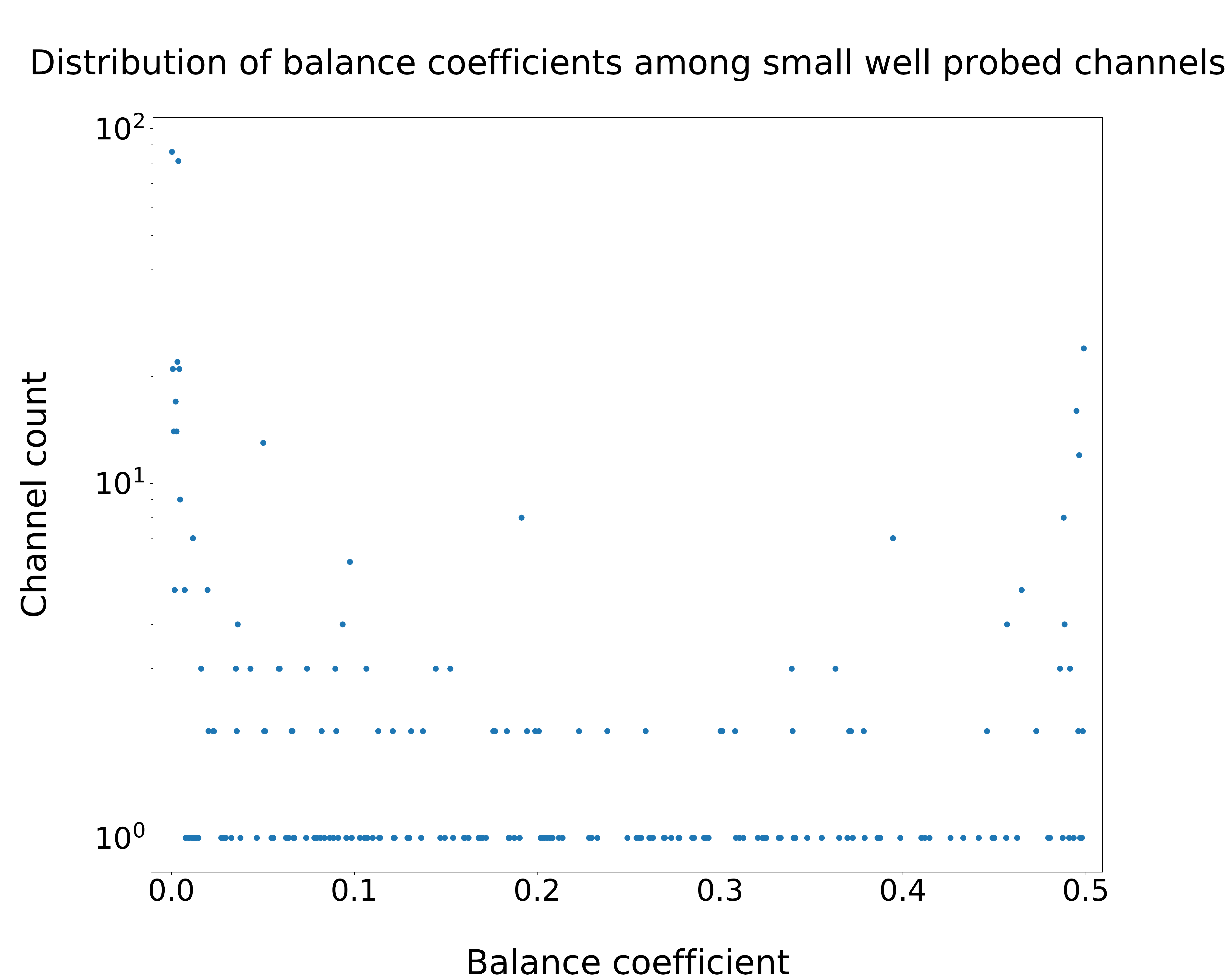}
  \caption{A large portion of channels is unbalanced (coefficient close to zero).}
  \label{fig:balance-coeff-histogram}
\end{figure}

Finally, we wanted to understand how often we send onions through each channel.
As Figure~\ref{fig:channel_frequency_distribution} shows, most of our routes go through a small number of channels.
It is partially explained by the fact that all routes by definition pass through one of our own entry channels.
We suggest that this distribution, which seems to follow the power law, is consistent with the properties of a small world network with a small diameter.

\begin{figure}[]
  \centering
  \includegraphics[scale=0.45]{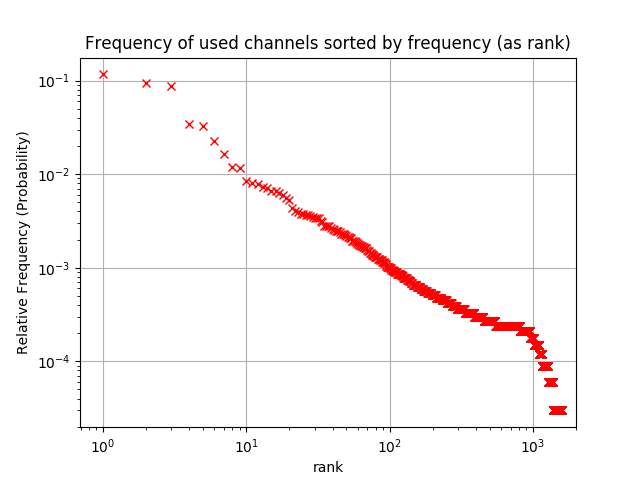}
  \caption{Distribution how often channels have been used in routes seems to follow a power law.}
  \label{fig:channel_frequency_distribution}
\end{figure}

\section{Discussion and countermeasures} \label{sec:discussion}
Our experiments show that channel balances can not be considered private data.
The current approach to managing balance information is neither used to optimize routing, not protected properly.
We envision two paths for LN development with either privacy or routing efficiency prioritized.
We outline potential countermeasures and leave their full development and evaluation for future work.

\subsection{Without protocol modification}
We envision one countermeasure that would partially mitigate the current probing technique.
This can be implemented as part of a node routing policy.
Note that all our payments used for probing fail (by design, either due to insufficient balance or due to intentionally wrong hash value). 
Intermediary nodes know if a payment they are a part of succeeds or fails.
Therefore, an intermediary node observing a flood of failing payments from the same channel may assume that this is a probing, especially if the amounts follow the binary search pattern.
Of course, such techniques can be tricked: an adversary can connect to Bob via Alice and make Bob think that Alice is malicious and is measuring his channel balance.

\subsection{Prioritizing privacy}
We argue that it is infeasible to reliably protect balance information of LN channels.
This conclusion comes from the following observations:
\begin{itemize}
	\item The sender knows whether the payment has failed or succeeded;
	\item The sender knows which channel has failed if the payment has failed.
\end{itemize}

However, we can modify the protocol to make the latter assumption to not hold.

\subsubsection{Merging error types}
To preserve privacy, we propose a modification to LN error handling.

The problem is that the two erring cases we distinguish differ in both the erring node and the error type.
In particular, if the target channel has insufficient capacity, the error is returned by the \textit{previous} node.
If the target channel has enough capacity, then the \textit{final recipient} reports incorrect payment details.

The goal of the proposed changes is to prevent the sender from knowing where the payment has failed.
In particular, we propose that each node in a route modifies the error it sends back as if it has originated from its own channel.
We also suggest merging the two error types ("incorrect or unknown payment details" and "temporary channel failure").
A similar countermeasure has already been implemented (see note about error types 16 and 17 in BOLT4~\cite{BOLT4}).
The drawback of this method is a decrease in payment reliability: as the sender does not know which channel has failed the payment, this channel can not be excluded in the subsequent route search.

\subsubsection{Additional loops}
Another potential countermeasure would be for intermediary nodes to add extra hops to the route.
Currently LN is source-routed: the route is chosen by the sender and enforced with onion routing.
Intermediary nodes can not change the sequence of nodes that a payment goes through (though they can choose a channel to the next node, if multiple channels are available).
If this scheme is modified, instead of offering an HTLC to Bob directly as prescribed by the initial sender, and intermediary node Alice would forward the payment to Bob through a random sub-route via Charlie and David.
This would blur the picture for the sender, as the sender wouldn't know which path the payment has really taken.

Possible drawbacks of this approach include a more complex fee structure and the requirement to substantially modify the onion routing protocol in the current LN.

\subsubsection{JIT routing}
JIT routing can be an effective countermeasure against channel probing attacks.

Just In Time Routing (JIT-routing) algorithm was originally proposed to improve payment reliability~\cite{pickhardt2019jit}.
In JIT routing, forwarding nodes send a circular payment to themselves if the balance of a channel is insufficient to forward a payment (channel rebalancing).
In the probing scenario, a JIT-supporting node will not send an error message back if it is lacking funds on the attacked (probed) channel.
Rather, it will interrupt the routing process, rebalance its channels, and continue forwarding the payment.
The attacker would interpret this as a signal that the target channel has sufficient funds (the same holds if the channel is probed from the other side).
The task now is to ensure that rebalancing is always be successful.
This can be established by ensuring that the max flow in the network is sufficient.

As JIT routing takes additional time, there is a possibility for a timing attack. 
However, the attacker has no control over where on the probed path JIT routing operations take place. 
We leave the evaluation of this proposal for future work.
We suggest to start by looking at the proactive channel rebalancing proposal~\cite{pickhardt2019imbalance} that can be reused to engage in various JIT-routing strategies. 

\subsection{Prioritizing routing efficiency}
If one concludes that it is infeasible to protect channel balances, then one may decide not to consider this information secret and utilize it to improve routing efficiency.

A common cause for routing failure in LN is the lack of balance information at the sender's node.
Routes are chosen under the assumption that a channel with capacity $c$ can route payments up to $c$, which is very optimistic.
The sender could filter out unsuitable routes if it knows the balances of all channels (as we do in Algorithm~\ref{alg:find-route}).
Broadcasting all intermediate balances to all nodes is unfeasible, as it would introduce a large networking overhead and structurally replicate the layer one (Bitcoin) gossip, where each transaction must reach every node.
This puts severe limitations on layer one scalability, which is the very problem LN is meant to solve.

We propose to develop reasonable means for nodes to share information about balances with other nodes, that does not overwhelm the network.
This information would help nodes make better path-finding decisions or recognize how to allocate their funds to newly created payment channels.

\subsubsection{Sharing balance data}
Instead of sharing all the balances, we propose adding an API call that would let the sender query the balance of a channel it wants to route a payment through.
In this scenario, a sender would create a preliminary route and query the nodes along this route on whether they have sufficient balance.
If some of them do not, the sender re-calculates the route until a suitable route is found.
This would eliminate or at least improve upon the current LN payment workflow, where a sender is receiving errors and re-sending a payment along multiple routes until it succeeds.
Nodes could come up with policies regarding balances, for instance, only reveal balances to trusted nodes, or only to nodes that pay a fee.
The ability of a node to reveal a channel balance for routing purposes may also be subject to negotiation between channel partners during channel establishment.
We should also think about ways to prevent abuse of this system.
We leave a more detailed analysis of this proposal and the evaluation of its efficiency to future work.

\subsection{Limitations}

Now we discuss the limitations of our approach and potential ways to improve it.

\subsubsection{Inactive channels}

We can only probe channels that allow routing through them (i.e., \textit{active} channels).
The routing function does not consider channels marked as not active.

\subsubsection{Private channels}

LN channels do not have to be announced.
In particular, casual users on mobile devices are not supposed to announce their channels.
These channels are called \textit{private}.
According to a recent study~\cite{BitMEXPrivateChannels}, $28\%$ of LN~channels are private.
Private channels are not prone to our probing methodology.

It may be possible to extend our technique to private channels by using on-chain heuristics to locate them.
In particular, each channel has a short ID composed of the number of the block, the transaction, and the UTXO index of the funding transaction.
Scanning the blockchain for the outputs of the corresponding form and cross-referencing it with the LN gossip may show private channels.
This attack has been reported and is in the process of being mitigated~\cite{LN675}.

\subsubsection{No route for the required amount}

We can not probe a route if we could not find a suitable route to a target channel.
In particular, we can not probe a high-capacity channel if it is connected to the rest of the network only through a low-capacity channel.
This limitation can be partially overcome by diverging from the series of probing amounts determined by binary search (where each new amount divides $[b_{min}, b_{max}]$ in half).
Instead, we could probe such channels with a smaller amount.
In the initial version of our algorithm, we did not do it for simplicity.
In any case, our probing capacity for the high-capacity target channel is limited by the capacity of (smaller) intermediary channels.

\subsubsection{Error types}

Our interpretation of error messages may not always be correct.
For instance, "temporary channel failure" can be caused by other failures apart from insufficient capacity.
Moreover, some of the channels respond with errors that we interpret as neither "insufficient capacity" nor "unknown preimage".
Interpreting errors more accurately may improve the results.
This can be achieved by inspecting the source code of LN~implementations and understanding exactly under which circumstances each error type is returned.

\subsubsection{Concurrency, large channels, and probing time}

Our algorithm prescribes that each next probing amount cuts $[b_{min}, b_{max}]$ in half.
For large channels, this is not always possible, as our current method only sends out one HTLC at a time to a specific route.

Lightning implementations impose limits on the maximal amount transferred in one HTLC.
This amount is approximately $0.043$~BTC.\footnote{4294967295 millisatoshis.}
The total channel capacity also has a soft limit of  $0.167$~BTC.\footnote{A soft limit means that it is a default restriction in the implementations, a node administrator can manually override it and open a larger channel is the counterparty agrees.}
This means that we can not perform the initial probe (with the amount $c / 2$) on channels larger than $0.086$~ BTC.
For such channels, we decrease the first probing amount to the maximum HTLC amount minus a safety margin.
\footnote{Trying to probe with exactly the maximum amount lead to errors, as we must also account for fees and possible safety margins at other nodes. In our code, we use the local maximum HTLC amount of $0.042$~BTC.}
If the capacity distribution is more unequal than around 1:3, we can continue probing, otherwise we admit that the true capacity distribution is in the interval which we can not probe with one HTLC.

However one can overcome this limitation by probing such channels with multiple concurrent HTLCs.
Our current algorithm is not concurrent, because we only control the sender, but not the receiver.
Therefore, if the receiver is live, an error returns quickly, and the capacity is unblocked.
To probe large channels, we need to block funds for longer.
This would involve a modified node acting as the recipient and deliberately delaying sending back an error, thus temporarily blocking funds along the route in an series of in-flight HTLCs.
This would allow us to have more precise lower bound estimates, in particular for large and distant channels.

The probing takes a noticeable amount of time (a few hours, depending on parameters).
Assuming that the usage of LN is not very high, we assume that the effect on our results is low.
But strictly speaking, our results do not constitute an instant snapshot of the network.
This limitation can also be overcome by concurrent probing.

However, adding concurrency is a non trivial task, as parallel probings may interfere with each other.
For example, if a channel with a balance $X$ is concurrently probed with amount $0.7X$ by two probing instances, it would return "insufficient capacity" to one of them, though it could have forwarded two probes of $0.7X$ each sequentially.
It is possible to parallelize the probing, but one has to ensure that the parallel probing agents do not interfere with each other.

\subsubsection{Parallel channels and non-strict forwarding}

Our method is based on an assumption that the payment is being forwarded through the channels as determined by the sender.
However, the LN specification only guarantees that the sequence of \textit{nodes} is followed.
We denote channels shared by the same pair of nodes as \textit{parallel}.
A forwarding node is free to choose a channel from all parallel channel to the next node in the route.
This provides more flexibility, as an intermediary node can choose an optimal channel based on balance restrictions.

In our setup, this means that we can not enforce our probes to be sent through a given channel.
The probe may be forwarded through a parallel channel.
This is also true for all channels along the route.
We accept this as a limitation of our approach.

Note also that while the LN specification (BOLT) allows non-strict forwarding, the c-lightning implementation that we use does not allow creating multiple channels to the same node.
The other two popular implementations, LND and Eclair, allow parallel channels.

\subsubsection{Hanging in-flight HTLCs}

Our method assumes that for each probe an error is returned quickly (within seconds, as Figure~\ref{fig:route_length_timings} shows).
If some intermediary hop does not return an error, we are left with an HTLC that we call \textit{hanging}.
Hanging HTLCs occupy the capacity of our channels, preventing us from probing with large amounts.
In that case, we must issue smaller probes and, as a result, obtain less information.
Hanging HTLCs are hard to get rid of: the protocol does not allow to cancel them unilaterally, and closing a channel involves long timeouts until the funds are available and can be put into a new channel.
This issue is at the core of the denial-of-service attack described in~\cite{mizrahi2020congestion}.

\subsection{Estimating the attack cost}

The attack requires moderate resources.
The attacker only needs to commit funds to open entry channels.
Entry channels' capacity determine the maximum amount of probes the attacker can send.
With our current approach, the maximum probing amount is the protocol's HTLC limit of $0.167$~BTC, or around $1500$~USD at the time of writing.
This is the minimal amount the attacker has to commit to theoretically be able to fully probe all "small" channels.
Our experience shows that it is better to open multiple channels to decrease the negative effect of hanging HTLCs.
In our experiments, we used five~entry channels.

An important feature of our attack is that since all the probing transactions fail, no fees have to be paid.
The only cost for the attacker is the time value of money.
If no hanging HLTCs are left after the attack, the attacker can close the entry channels collaboratively and withdraw the committed funds immediately.
If some HTLCs are hanging, or if the attacker's channel partners are offline or unwilling to cooperate on channel closure, the attacker would have to wait for the agreed upon timeout (usually on the order of days).
In any case, no funds committed to the attack are irrevocably lost.

\section{Related work} \label{sec:related-work}
Herrera{-}Joancomart{\'{\i}} et al.~\cite{herrera2019difficulty} propose a balance probing attack.
Their approach is similar to ours, but only allows the attacker to probe the channels immediately adjacent to the node it has opened a channel with.
In contrast, our algorithm only requires opening a few entry channels for probing the whole network using arbitrarily long routes.\footnote{Up to the protocol limitation of 20~hops.}



Conoscenti et al.~\cite{conoscenti2019hubs} analyzed the influence of hubs on LN and proposed a channel rebalancing algorithm.

B{\'{e}}res et at.~\cite{beres2019cryptoeconomic} provided a cryptoeconomic analysis of LN and argues that despite onion routing privacy of payments can be breached due to short routes and strong statistical hints.

Tang et al.~\cite{tang2019privacyutility} showed that introducing noise in payment channel balances does not bring substantial improvements in privacy without hurting routing efficiency.

Mizrahi and Zohar~\cite{mizrahi2020congestion} describe a denial-of-service attack on LN based on creating and not redeeming HTLCs along long routes.


\section{Conclusion} \label{sec:conclusion}
LN is a promising off-chain scaling solution for Bitcoin.
One of the major issues LN is yet to solve is routing efficiency: senders must choose routes with no information on balance distributions, causing payment failures.
Balance information can not be shares universally because of scalability and privacy reasons.
However, despite the intuitive impression that LN provides a high level of privacy by using onion routing and not broadcasting transactions, we argue that LN channel balances are not well protected.
Our experiments show that a low-resource attacker can probe the balances of the majority of live and active channels, revealing their balances and potentially tracking the flow of value through selected channels in near real time.
We implement and evaluate our technique on the Bitcoin testnet, successfully probing a large portion of channels.
We consider the tradeoff between balance privacy and routing efficiency, and describe multiple directions for future work that would evaluate the ways to find the right balance between the two.



\bibliography{balanceDistribution}
\bibliographystyle{plain}

\end{document}